\documentclass[aps,prd]{revtex4}
\usepackage{subfigure,graphics} 
\usepackage{flafter}
\usepackage{multirow}
\pdfoutput=1
\begin{document} 

\title{
Black holes with current loops revisited} 
\author{Ian G. Moss}
\email{ian.moss@ncl.ac.uk}
\affiliation{School of Mathematics and Statistics, Newcastle University, NE1 7RU, UK}

\date{\today}


\begin{abstract}
The electromagnetic field around a Kerr black hole inside a current loop is sometimes
used as the basis of a toy model for discussing the properties of particle orbits near astrophysical
black holes. The motivation for the present paper is to correct the published solution to Maxwell's
equations with a charged current loop. Dipole approximations and closed-form expressions in the
extreme Kerr limit are also presented. Using the corrected solution, it turns out that imposing a
vanishing electromotive force produces a loop with a potential which is finite everywhere outside
the black hole. Ring solutions can be combined into solutions with multiple rings or current discs.
\end{abstract}
\pacs{PACS number(s): }

\maketitle
\section{introduction}

Attempts have been made over the years to explain the generation of untrarelativistic particles
through physical effects around black holes located inside accretion disks
(see \cite{Punsley:2001} for a review). In some situations, large electric potentials can be
produced by dynamo action if the black hole is rotating in a magnetic field generated in the
accretion disk. In the original Blandford-Znajek model,  for example, the induced  electric fields
drive currents and act as the source of energy fluxes \cite{Blandford:1977,Thorne:1986iy}.

In the context of some of these models, it becomes important to study particle
orbits and particle interactions in electric and magnetic fields close to the black hole. These
studies are increasingly being replaced by magnetohydrodynamics (MHD), but the particle picture
is needed when a fluid description is not appropriate.  The study of circular orbits, for example,
gives the location for the inner edge of the accretion disk for free charges
\cite{Prasanna:1978,Prasanna:1980,Aliev:2002nw}.
In the presence of the magnetic field, the equations governing the charged particle orbits are no
longer seperable, and the particle motion can be come quite complicated outside of the equatorial
plane \cite{Sengupta:1997na,Takahashi:2008zh}. This may well have an effect on issues such as
charge separation and the validity of fluid approximations.

The large potential around the black hole implies that Penrose processes can be an important source
of untrarelativistic particles \cite{Wagh:1985}. Familiar limits on the efficiency of the Penrose
process \cite{Bardeen:1972,Wald:1974b} are not valid for charged particles, and the particles
emerging from the ergoregion can have energies close to the available electric potential
\cite{Wagh:1989}. For example, magnetic fields of $10^4$ Gauss around $10^9$ solar mass black holes
can produce positrons and electrons with energies as high as $10^{20}{\rm eV}$ through reactions
such as $\gamma\gamma\to e^+e^-$ inside the ergoregion.

All of the discussions of particle motion mentioned above have used the field from a current loop,
or its dipole limit, as a toy model to represent the magnetic field. The solution to Maxwell's
equations for the uncharged current loop was obtained a long time ago by Chitre and Vishveshwara
\cite{Chitre:1975}, and generalised to a charged current loop by Petterson \cite{Petterson:1975}. 
The general class of stationary solutions was analysed by Bi\v{c}{}\'ak and Dvo\v{r}{}\'ak
\cite{Bicak:1976}, who also considered some special cases including current loops, dipoles and point
charges. Surprisingly, the result of Petterson is not consistent with the other two papers,
which is unfortunate because this is the only published solution which solves for the vector
potentials that are needed to study the particle dynamics.

In order to explain the differences further, consider the general vacuum axisymmetric solution 
to Maxwell's equations, with a set of four arbitrary constants for each angular mode. 
These constants can be related to 
the current and charge by matching vacuum solutions inside and outside the current loop. The 
inconsistency between the three papers mentioned above is due to an error in Ref 
\cite{Petterson:1975} when matching the solutions. (There is also a minor typographical
error in the constants of Ref. \cite{Petterson:1975}). Correcting the values of these constants
leads to physical consequences, for example the potential at the
horizon of the black hole changes and the zero EMF condition used in some astrophysical scenarios 
occurs for a different value of the black hole charge. 
The errors in Ref. \cite{Petterson:1975}
have subsequently been reproduced in standard texts \cite{Punsley:2001}. The motivation for the
present paper is to present the correct electromagnetic potentials for a charged current
loop.
 
The current loop solutions are appropriate when there are no free charges surrounding the black hole
and there is no radial current flow. In real situations the region
surrounding the black hole may contain particle-antiparticle pairs which conduct current freely.
This type of model could be built up by superimposing radial currents with the ring solutions.
The simplest possibility is to have radial current flowing out along the equatorial plane and then
back along the symmetry axis, as in the toy model by Li \cite{Li:2000}.

There are problems with the simple Blandford-Znajek model, for example it is likely that most
magnetic field lines pass through the accretion disk \cite{Punsly:1981} and the disk is then
subject to a variety of MHD instabilities \cite{Balbus:1991ay}. On the other hand, a similar model
may still hold in systems with relatively low accretion rates where the inner parts of the disk are
effectively Keplerian orbits \cite{Li:2000fn}. These issues are not considered further here, and the
results below can be used when the electromagnetic fields generated by the accretion disk can be 
replaced by the fields generated by current loops.

\section{Maxwell fields in the Kerr background}

The general solution to Maxwell's equations on a Kerr background is described in the textbook by
Chandrasekhar \cite{Chandra:1992}. In principle, we ought to solve the full Einstein-Maxell
equations, but the gravitational back-reaction of the magnetic field can be ignored when $BM<<1$ in
gravitational units, or
\begin{equation}
B<<2\times 10^{10}\left({M\over 10^9 M_\cdot}\right)\,\,{\rm G}
\end{equation}
in Gauss. In typical astrophysical situations, the electromagnetic field strengths are
likely to be insufficiently large to have any significant effect and the Kerr metric for a rotating
black  hole can be used consistently.

The Kerr metric in Boyer-Lindquist coordinates $(t,r,\theta,\phi)$ is given by
\begin{equation}
ds^2=-{\Delta\over\Sigma}\omega_1^2+{\sin^2\theta\over\Sigma}\omega_2^2
+{\Sigma\over\Delta}dr^2+\Sigma d\theta^2
\end{equation}
where
\begin{eqnarray}
\omega_1=dt-a\sin\theta d\phi,&&\omega_2=(r^2+a^2)d\phi-a dt\\
\Sigma=r^2+a^2\cos^2\theta,&&\Delta=r^2+a^2-2mr
\end{eqnarray}
The Maxwell field strength $F_{\mu\nu}$
is represented by three complex scalar quantities
\begin{eqnarray}
\phi_0&=&F_{\mu\nu}l^\mu m^\nu\\
\phi_1&=&F_{\mu\nu}(l^\mu n^\nu+m^{\mu*}m^\nu)/2\\
\phi_0&=&F_{\mu\nu}m^{\mu*} n^\nu
\end{eqnarray}
where the Newman-Penrose tetrad vectors are
\begin{eqnarray}
l^\mu&=&{1\over\Delta}\left(r^2+a^2,\Delta,0,a\right)\\
n^\nu&=&{1\over 2\Sigma}\left(r^2+a^2,-\Delta,0,a\right)\\
m^\mu&=&{1\over\bar\rho\sqrt{2}}\left(ia\sin\theta,0,1,i{\rm cosec}\theta\right)
\end{eqnarray}
and $\bar\rho=r+i\cos\theta$.

The stationary axisymmetric solutions to Maxwell's equations with vanishing sources are determined by solving the
Teukolsky equation
\begin{equation}
-\Delta\Psi_{,rr}-{\rm cosec}\theta(\sin\theta\Psi_{,\theta})_{,\theta}+(\cot^2\theta+1)\Psi=0
\end{equation}
where commas denote derivatives and the field $\Psi=-\Delta\phi_0$.
The general solution for $\phi_0$ can be expressed in terms of angular mode functions 
$P_l{}^1(\cos\theta)$,
\begin{equation}
\phi_0={1\over\sqrt{2}}\sum_{l=1}^\infty
\left(\alpha_lP_{l,r}(u)P_l{}^1(\cos\theta)+\beta_lQ_{l,r}(u)P_l{}^1(\cos\theta)\right)\label{sfsol}
\end{equation}
The coefficients $\alpha_l$ and $\beta_l$ are complex constants and $P_l(u)$ and $Q_l(u)$ are
Legendre functions of $u=(r-M)/(r_+-M)$, where $r_+$ is the radial coordinate of the event horizon.
The solution can be checked using the identity
\begin{equation}
(\Delta P_{l,r}(u))_{,r}=l(l+1)P_l(u).
\end{equation}
The terms involving $P_l(u)$ are regular at
the horizon and the terms involving $Q_l(u)$ are regular at spatial infinity. Typically, we
have different inner and outer solutions separated by a region with charges and currents.

The reconstruction of the gauge potentials given in ref \cite{Chandra:1992} is incomplete, but with 
a little extra work it is possible to extend the method used there to rederive the inner solution
given by Petterson \cite{Petterson:1975}
\begin{eqnarray}
A_t&=&\alpha_t-{rQ\over \Sigma}-\sum_{l=1}^\infty\left\{
{\Delta\over\Sigma}P_{l,r}(u)P_l(\cos\theta)(r\alpha_l^r-a\cos\theta\alpha_l^i)
+{a\over\Sigma}P_l(u)\sin\theta P_l{}^1(\cos\theta)(a\cos\theta\alpha_l^r+r\alpha_l^i)\right\}\\
A_\phi&=&{arQ\over \Sigma}\sin^2\theta+\sum_{l=1}^\infty\left\{
{a\Delta\over\Sigma}P_{l,r}(u)P_l(\cos\theta)\sin^2(\theta)
(r\alpha_l^r-a\cos\theta\alpha_l^i)\right.\\
&&\left.+{r^2+a^2\over\Sigma}P_l(u)
\sin\theta P_l{}^1(\cos\theta)(a\cos\theta\alpha_l^r+r\alpha_l^i)
-{\Delta\over l(l+1)}P_{l,r}(u)P_l{}^1(\cos\theta)\sin\theta\alpha_l^i   \right\}
\end{eqnarray}
where $Q$ is the black hole charge, $\alpha_t$ is a constant and $\alpha_l=\alpha_l^i+i\alpha_l^r$.
The possibility of a constant term in $A_\phi$ is excluded by the regularity of the vector potential
along the axis of rotation.

The outer solution uses the alternate set of Legendre functions, but it is also important to account
for the charge of the source region $q$. Since this will be important later, it is worth noting the
full outer solution,
\begin{eqnarray}
A_t&=&-{r(Q+q)\over \Sigma}-\sum_{l=1}^\infty\left\{
{\Delta\over\Sigma}Q_{l,r}(u)P_l(\cos\theta)(r\beta_l^r-a\cos\theta\beta_l^i)
+{a\over\Sigma}Q_l(u)\sin\theta P_l{}^1(\cos\theta)(a\cos\theta\beta_l^r+r\beta_l^i)\right\}\\
A_\phi&=&{ar(Q+q)\over \Sigma}\sin^2\theta+\sum_{l=1}^\infty\left\{
{a\Delta\over\Sigma}Q_{l,r}(u)P_l(\cos\theta)\sin^2(\theta)
(r\beta_l^r-a\cos\theta\beta_l^i)\right.\label{atgen}\\
&&\left.+{r^2+a^2\over\Sigma}Q_l(u)
\sin\theta P_l{}^1(\cos\theta)(a\cos\theta\beta_l^r+r\beta_l^i)
-{\Delta\over l(l+1)}Q_{l,r}(u)P_l{}^1(\cos\theta)\sin\theta\beta_l^i   \right\}
\end{eqnarray}
The gauge has been chosen so that $A_t$ vanishes at spatial infinity.

\section{Charged current loop}

The field due to the charged current loop is obtained by matching the inner
and outer solutions at the position of the loop. The safest way to do this is to use
the Maxwell scalars $\phi_0$ or $\phi_2$, because these are readily expressed
in terms of orthogonal polynomials in $\cos\theta$. Matching $\phi_1$, as was attempted in Ref
\cite{Petterson:1975}, leads to errors because $\phi_1$ contains both $P_l(\cos\theta)$ and 
$P_l{}^1(\cos\theta)$, and these are not orthogonal functions.

The current density for the charged current loop in the Boyer-Lindquist coordinates is given by
\cite{Bicak:1976}
\begin{equation}
J^\mu=\left( {q\over 2\pi r_0^2}\delta(r-r_0)\delta(\cos\theta),0,0,
{I\over r_0^2}\delta(r-r_0)\delta(\cos\theta)\right).\label{current}
\end{equation}
The parameter $q$ is equal to the total charge of the loop, as defined by an integration of the
charge density over a surface $S$ of constant time,
\begin{equation}
\int_S J^\mu d\Sigma_\mu=q.
\end{equation}
Gauss' law, applied to a large sphere, implies that this is the same charge which appears
in the potential (\ref{atgen}). 

The relationship between the Maxwell scalars and the source is obtained through the
corresponding Teukolsky equation,
\begin{equation}
-\Delta\Psi_{,rr}-{\rm cosec}\theta(\sin\theta\Psi_{,\theta})_{,\theta}+(\cot^2\theta+1)\Psi=
4\pi J
\end{equation}
The source function is related to the current density by
\begin{equation}
J=\Sigma(\Delta+\mu)\Sigma^{-1}J_{m^*}-\Sigma(\delta^*+\pi-\tau^*)\Sigma^{-1}J_n,\label{defj}
\end{equation}
where $\Delta$ and $\delta^*$ are directional derivatives along $n$ and $m^*$ respectively, and
\begin{equation}
\mu=-{\Delta\over2\Sigma\bar\rho^*},\quad \pi={ia\sin\theta\over\bar\rho^{*2}\sqrt{2}},\quad
\tau=-{ia\sin\theta\over\Sigma\sqrt{2}}.
\end{equation}
For axisymmetric sources, we may express the fields in angular modes with components $R_l$, where
\begin{equation}
R_l=\int_{-1}^1\Psi P_l{}^1(\cos\theta)d\cos\theta,\label{rlint}
\end{equation}
and similarly for $J_l$. These components satisfy
\begin{equation}
R_{l,rr}-{l(l+1)\over \Delta}R_l=-8\pi {J_l\over\Delta}.\label{rleq}
\end{equation}
In the source-free regions, $\phi_0$ is given by Eq. (\ref{sfsol}), and the integration
(\ref{rlint}) yields
\begin{eqnarray}
R_l&=&{-\Delta\over\sqrt{2}}{2l(l+1)\over 2l+1}\alpha_lP_{l,r}(u)\hbox{ inner region}\\
&=&{-\Delta\over\sqrt{2}}{2l(l+1)\over 2l+1}\beta_lQ_{l,r}(u)\hbox{ outer region}
\end{eqnarray}

When the current density (\ref{current}) is substituted into eq. (\ref{defj}), the differential
equation (\ref{rleq})  for $ R_l$  becomes
\begin{equation}
R_{l,rr}-{l(l+1)\over \Delta}R_l=F(r)\delta(r-r_0)_{,r}+G(r)\delta(r-r_0),\label{rldeq}
\end{equation}
where
\begin{eqnarray}
F(r)&=&-{i\over \sqrt{2}}{4\pi I(r_0^2+a^2)-2aq\over r_0}P_l{}^1(0)\label{Feq}\\
G(r)&=&-{i\over \sqrt{2}}4\pi I P_l{}^1(0)+{l(l+1)\over \sqrt{2}}{4\pi aI-2q\over r_0}P_l(0)
\label{Geq}
\end{eqnarray}
Integrating Eq. (\ref{rldeq}) through the delta function gives
\begin{eqnarray}
&&-{\Delta\over\sqrt{2}}{2l(l+1)\over 2l+1}\left(\beta_lQ_{l,r}-\alpha_lP_{l,r}\right)=F(r_0)
\label{fbc}\\
&&-{\Delta\over\sqrt{2}}{2l(l+1)\over 2l+1}\left(\beta_lQ_{l}-\alpha_lP_{l}\right)=
{\Delta\over l(l+1)}G(r_0).
\end{eqnarray}
Combining these equation gives the solution for $\beta_l$,
\begin{equation}
\beta_l={l+1/2\over l(l+1)}{1\over w}\left\{
{\sqrt{2} G(r_0)\over l(l+1)}P_{l,r}(u_0)-{\sqrt{2}F(r_0)\over\Delta(r_0)} P_l(u_0)\right\},
\end{equation}
where $w$ is the Wronskian,
\begin{equation}
w=P_l(u_0)Q_{l,r}(u_0)-Q_l(u_0)P_{l,r}(u_0)={r_+-M\over \Delta(r_0)}.
\end{equation}
The corresponding solution for $\alpha_l$ is obtained by replacing $P_l$ by $Q_l$. The real and
imaginary parts of $\beta_l$ can be identified by using Eqs. (\ref{Feq}) and (\ref{Geq}),
\begin{eqnarray}
\beta_l^r&=&{l+1/2\over l(l+1)}{1\over w}{4\pi aI-2q\over r_0}
P_{l,r}(u_0)P_l(0)\\
\beta_l^i&=&{l+1/2\over l(l+1)}{1\over w}\left\{
{4\pi I(r_0^2+a^2)-2aq\over r_0\Delta(r_0)}P_l(u_0)-{4\pi I\over l(l+1)} P_{l,r}(u_0)\right\}
P_l{}^1(0)
\end{eqnarray}
Similarly, for $\alpha_l$,
\begin{eqnarray}
\alpha_l^r&=&{l+1/2\over l(l+1)}{1\over w}{4\pi aI-2q\over r_0}
Q_{l,r}(u_0)P_l(0)\\
\alpha_l^i&=&{l+1/2\over l(l+1)}{1\over w}\left\{
{4\pi I(r_0^2+a^2)-2aq\over r_0\Delta(r_0)}Q_l(u_0)-{4\pi I\over l(l+1)} Q_{l,r}(u_0)\right\}
P_l{}^1(0)
\end{eqnarray}
These results are consistent with Chitre and Vishveshwara \cite{Chitre:1975} in the $q=0$ limit,
allowing for the different definition of $I$. The real parts differ from Petterson
\cite{Petterson:1974,Petterson:1975} by a factor of $r_0$, but the imaginary parts are totally
different.

The constant term $\alpha_t$ remains to be evaluated, and for this we use the field along
the axis of rotation,
\begin{eqnarray}
A_t&=&-{\Delta\over r^2+a^2}\sum_{l=1}^\infty\left(
r\alpha_l^r-a\alpha_l^i\right)P_{l,r}(u)-{rQ\over r^2+a^2}+\alpha_t\label{ataxis}\\
A_t&=&-{\Delta\over r^2+a^2}\sum_{l=1}^\infty\left(
r\beta_l^r-a\beta_l^i\right)Q_{l,r}(u)-{r(Q+q)\over r^2+a^2}
\end{eqnarray}
For $\alpha_t$, we require continuity of $A_t$ at $r=r_0$. From Eq. (\ref{fbc}), we get
\begin{equation}
\alpha_t=-{r_0q\over r_0^2+a^2}-{\sqrt{2}a F^i\over r_0^2+a^2}\sum_{l=1}^\infty{l+1/2\over l(l+1)}.
\end{equation}
Evaluating the sum and using Eq. (\ref{Feq}) gives
\begin{equation}
\alpha_t={2\pi Ia-q\over r_0}.\label{alphat}
\end{equation}
Note that the term $\alpha_t$ represents the value of the potential at the horizon of an uncharged
black hole.

\begin{center}
\begin{figure}[htb]
\scalebox{1.0}{\includegraphics{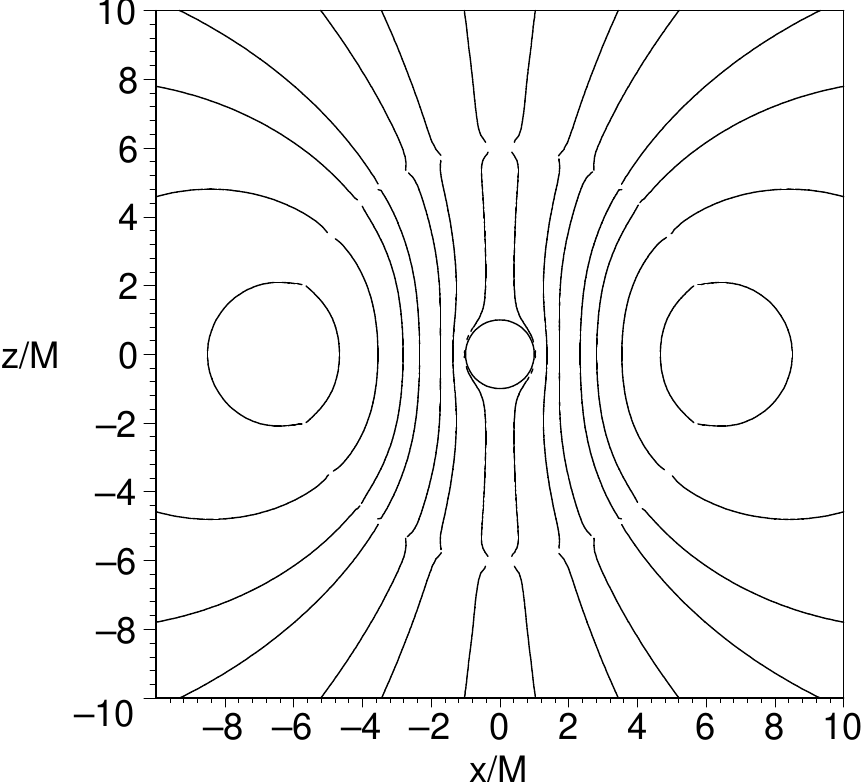}}
\scalebox{1.0}{\includegraphics{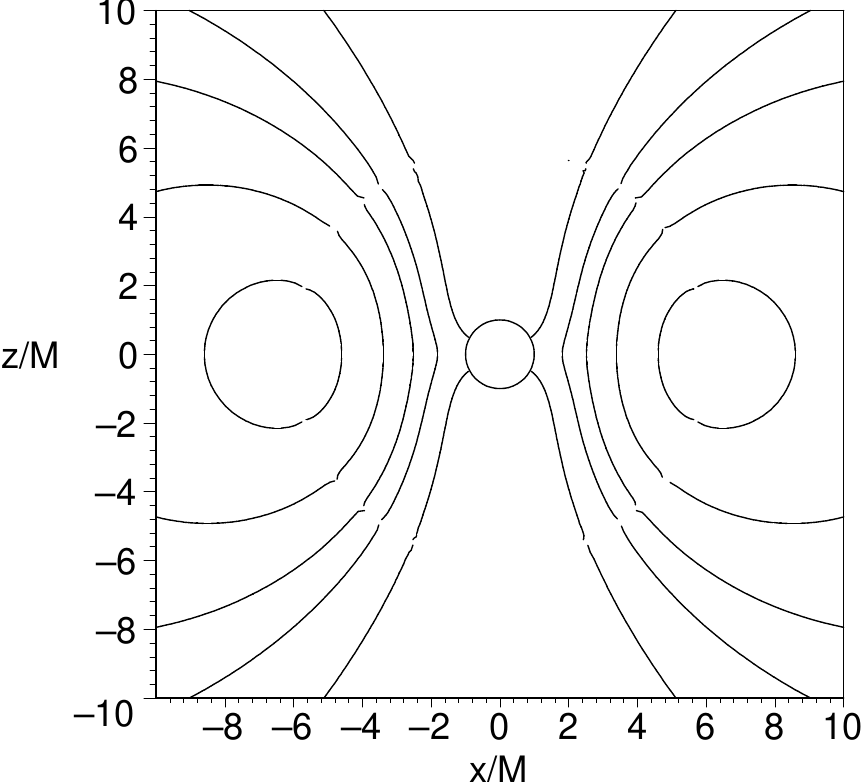}}
\caption{The magnetic field lines are shown in a vertical plane through the axis for an extreme Kerr
black hole with an equatorial current loop at $r=6M$ (left), and for an extreme Kerr-Newman black
hole (right).
The mode sum has been truncated at $l=16$, and the small irregularities ar $r=6M$ are due to this
truncation. }
\label{fig1}
\end{figure}
\end{center}

The magnetic field lines (in any synchronous frame) can be traced by the contours of constant
$A_\phi$, and a couple of examples
are shown in Fig. \ref{fig1}. The plot on the left shows the field lines in a vertical plane
through the axis for an extreme Kerr black hole with a current loop at $r=6M$, and the plot on the
right shows the field lines for an extreme Kerr-Newman black hole (with a small charge).
The former case is an example where the black hole expels the magnetic flux at the horizon. 
This black hole version of the Meisner effect has been noted before 
\cite{Bicak:1980,Bicak:1985,Chamblin:1998qm}.
The plots shown here demonstrate clearly that extreme black holes can have magnetic flux 
across the horizon when they are charged. In fact,
$A_\phi=Q/2$ at the black hole equator and $A_\phi=0$ at the north pole.

\section{Charge accretion}

The physical interpretation of the horizon potential is that the rotation 
of the black hole combined with the magnetic field of the loop generates a large electromotive force
(EMF). In an astrophysical context, as first pointed out by Wald \cite{Wald:1974}, free charges can
migrate along the magnetic field lines to cancel the EMF, resulting in a charged black hole.

If the total system of black hole plus current loop is to remain neutral, then charges should also
migrate to the current loop. We shall refer to the resulting solution as the zero-EMF solution with
vanishing total charge. The  zero-EMF and vanishing total charge conditions where analysed in
Ref. \cite{Petterson:1975}, but these results used an incorrect form for the electromagnetic
potential. The charge on the current loop was corrected by Linet \cite{Linet:1977,Linet:1979}, but
without obtaining the vector potential. (Note that Linet uses a different definition of the current
$I$.) The full solution is given below for the first time.

\subsection{Zero-EMF solution with vanishing total charge}

\begin{center}
\begin{figure}[htb]
\scalebox{0.45}{\includegraphics{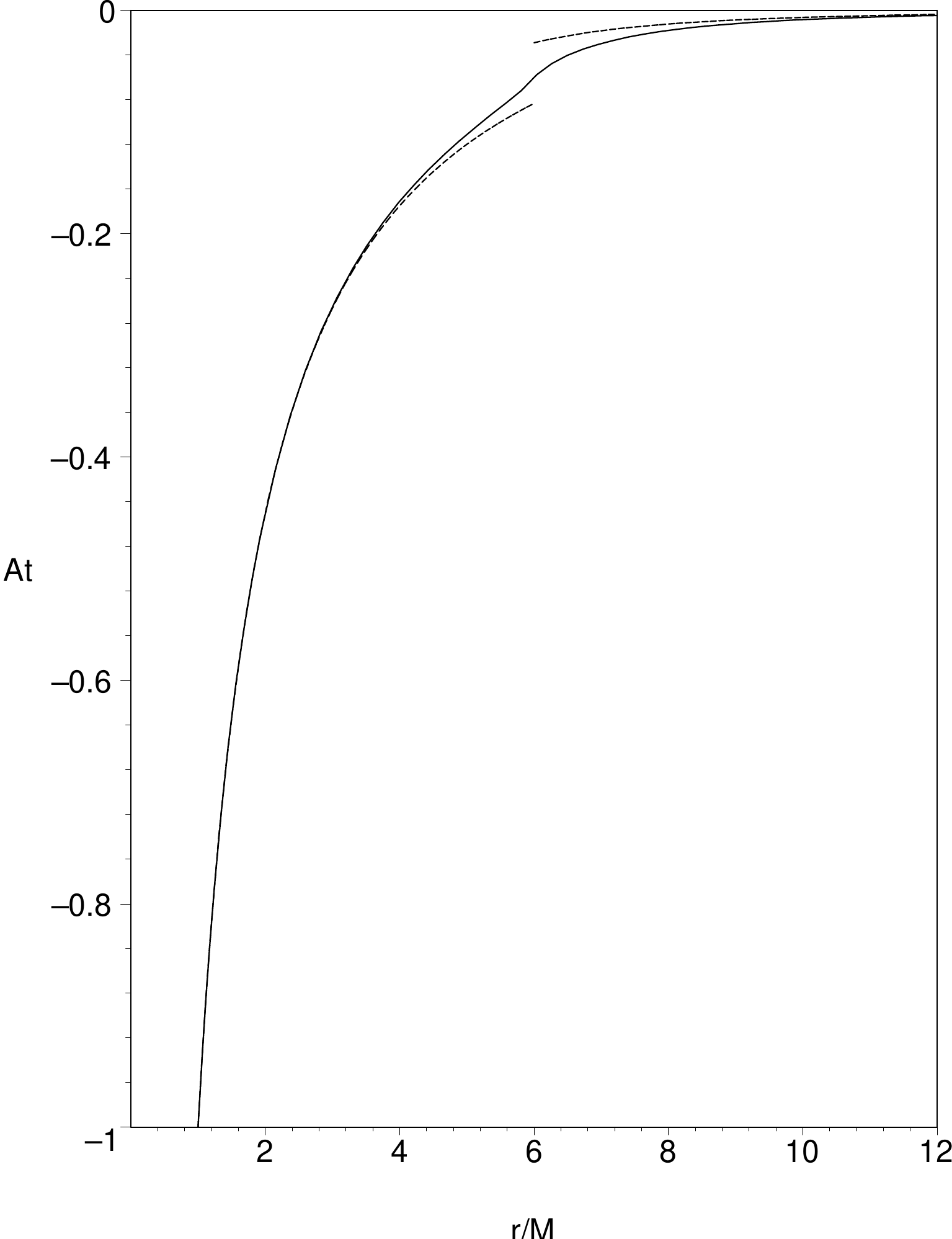}}
\scalebox{0.45}{\includegraphics{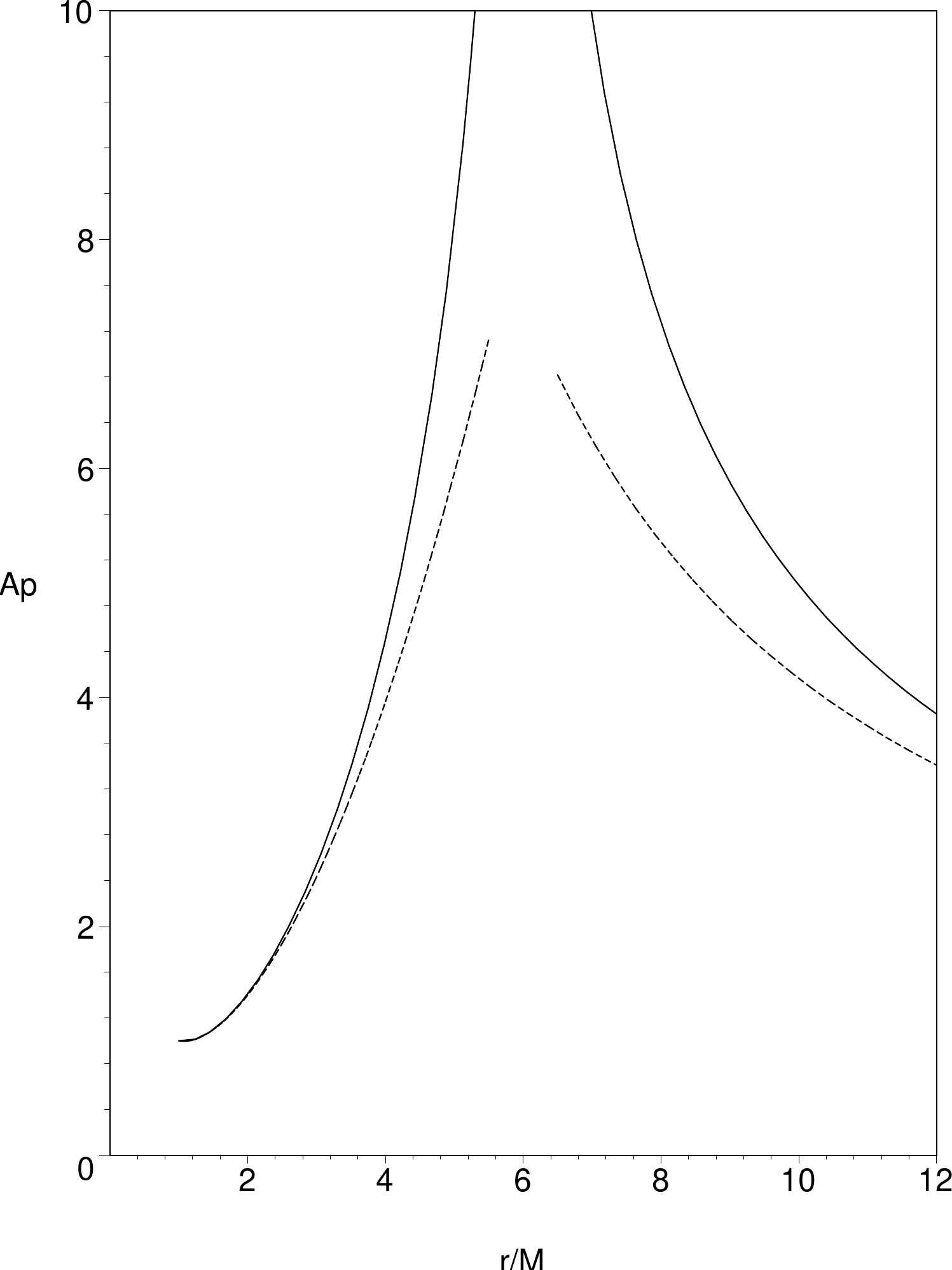}}
\caption{The vector potential solutions in the equatorial plane
for a loops with radius $6M$ around an extreme Kerr black hole. 
The dipole solution for a field of strength $B$ is also shown. The fields
are scaled by the modulus of the horizon value.}
\label{fig2}
\end{figure}
\end{center}

It will be sufficient to consider the electrostatic potential (\ref{ataxis}) along the axis of
rotation. The gauge has been chosen so that the potential vanishes at infinity, and so the total EMF
is given by the potential $-A_t$ at the horizon $r_+$. It vanishes when
\begin{equation}
\alpha_t={Q\over 2M}.\label{emf}
\end{equation}
The quantity $\alpha_t$ was evaluated in (\ref{alphat}). If we require vanishing total charge then
$q=-Q$ and the loop has a charge
\begin{equation}
q=-{4\pi MIa\over r_0-2M}\label{totq}.
\end{equation}
Examples of the vector potentials for the zero-EMF solution are shown in Fig. \ref{fig2}.

It is interesting to note that there is a lower limit to the radius of the loop in the
zero-EMF solution when $a>M/2$. Suppose that the loop contains $N_+$ particles of 
charge $e$ and $N_-$ particles of charge $-e$,
so that
\begin{equation}
q=(N_+-N_-)e.
\end{equation}
The current is bounded by having all of the charges moving at close to the speed of light,
hence \footnote{A more careful analysis using the geodesic equations for the motion of the charges
gives a stronger limit.}
\begin{equation}
|I|<{(N_++N_-)e\over 2\pi r_0}.\label{ilim}
\end{equation}  
The charge $q=-Q$ is related to the current by (\ref{totq}), and the limit (\ref{ilim}) implies that
\begin{equation}
r_0>M+(2Ma-M^2)^{1/2}
\end{equation}
when $a>M/2$. This is larger than the radius of the minimal circular stable orbit for
black holes close to the extremal limit.

\subsection{Finitenes of the potential}\label{sec}

It is natural for the electromagnetic fields to diverge in the vicinity of a charged current loop,
and it is remarkable that the electrostatic potential plotted in Fig. \ref{fig2} appears to be
finite. This turns out to be a general feature of the zero-EMF solution, where both the
electrostatic potential and the electric field remain finite at the loop
in the Boyer-Lindquist coordinate system. As a consequence, loops of this type are useful
when combined into an equatorial disc of constant potential.

Consider the leading terms in the multipole expansion (\ref{atgen}) which give a divergence
as $r$ approaches $r_0$ in the equatorial plane,
\begin{equation}
A_t\sim - {\Delta(r_0)^2f\over r_0^2(r_+-M)}
\sum_{l=1}^\infty{l+1/2\over l(l+1)}Q_{l,r}(u_0)P_{l,r}(u_0)P_l(0)^2x^l
-{ag\over r_0^2(r_+-M)}
\sum_{l=1}^\infty{l+1/2\over l(l+1)}Q_{l}(u_0)P_{l}(u_0)P_l{}^1(0)^2x^l
\end{equation}
where $f=4\pi Ia-2q$, $g=4\pi I(r_0^2+a^2)-2aq$ and $x$ is a dummy variable which has $x\to 1$
as $r\to r_0$. Standard formulae for the large $l$ limits
of the Legendre functions give,
\begin{eqnarray}
P_l(u_0)Q_l(u_0)&\sim&{1\over 2l+1}(r_+-M)\Delta(r_0)^{-1/2}\\
P_{l,r}(u_0)Q_{l,r}(u_0)&\sim&{l(l+1)\over2l+1}(r_+-M)\Delta(r_0)^{-3/2}
\end{eqnarray}
The divergent terms in the sum are then
\begin{equation}
A_t\sim -{\Delta(r_0)^{1/2}f\over 2r_0^2}
\sum_{l=1}^\infty P_l(0)^2x^l
-{ag\over 2\Delta(r_0)^{1/2}r_0^2}
\sum_{l=1}^\infty {P_l{}^1(0)^2\over l(l+1)} x^l
\end{equation}
The sums can be expressed in terms of hypergeometric functions,
\begin{eqnarray}
\sum_{l=0}^\infty P_l(0)^2x^l&=&F(1/2,1/2;1;x^2)\\
\sum_{l=1}^\infty {P_l{}^1(0)^2\over l(l+1)} x^l&=&xF(1/2,1/2;1;x^2)-{x\over 2}F(1/2,1/2;2;x^2).
\end{eqnarray}
The hypergeometric function $F(1/2,1/2;1;x^2)$ diverges as $x\to 1$. The potential
is convergent if, and only if, the coefficient of this hypergeometric function vanishes, and that
happens when
\begin{equation}
q=-{4\pi I Ma\over r_0-2M}
\end{equation}
Consequently, in view of Eq. (\ref{totq}), the zero-EMF solution with vanishing total charge has a
finite potential.

In an inertial frame, Gauss' law implies that the electric field of a charged loop diverges,
and so a loop can only have a finite potential if there exists an inertial frame
in which the charge of the loop vanishes. The appropriate frame is the one with orthogonal basis
vectors
\begin{eqnarray}
e_t{}'&=&(1,0,0,0)\\
e_\phi{}'&=&\left({-2Mar\sin^2\theta\over \Sigma-2Mr},0,0,1\right)
\end{eqnarray}  
This is the stationary frame, rather than the usual co-rotating frame of the Kerr black hole.
We conclude that the zero-EMF solution with vanishing total charge also has vanishing
local charge in the stationary frame.  

\section{Special limits}

The charged ring solution is likely to be most useful for giving the electromagnitic fields in
vacuum regions close to the hole. In these regions a dipole truncation of the electromagnetic field
can be a used as an  approximation to the full field. The solution is especially useful for
locating the innermost stable circular  orbits \cite{Aliev:2002nw} and for examining the production
of energetic particles by the Penrose process. 

The effective ergosphere in which the Penrose process can take place extends far
out  along the equatorial plane with a relatively narrow height along the rotation axis
\cite{Aliev:1988wy}. Therefore, the fields in the equatorial plane can be used for studies of both
stable orbits and the Penrose process. For a current loop around an extreme Kerr solution it is
possible to obtain the vector potential in the equatorial plane in a useful closed form given
below.

\subsection{Dipole and quadrupole truncations}

Starting with the inner solution, the Coulomb and Dipole terms combine into
\begin{eqnarray}
A_t&=&-{aMBr\over\Sigma}\sin^2\theta+(Q-2aMB)\left({1\over 2M}-{r\over\Sigma}\right)\\
A_\phi&=&{BA\over\Sigma}\sin^2\theta+(Q-2aMB){ar\over\Sigma}\sin^2\theta
\end{eqnarray} 
where $A=(r^2+a^2)^2-\Delta a^2\sin^2\theta$ and
\begin{equation}
B={\alpha_1^i\over r_+-M}.
\end{equation}
The $B$-dependent terms in each field component represent the fields of a black hole immersed in a
constant magnetic field of strenght $B$ \cite{Wald:1974}. Fig. \ref{fig2} shows the dipole
approximation for the zero-EMF solution with an extremal hole and a loop of radius $6M$. The
approximation agrees remarkably well for the $A_t$ component and reasonably well for $A_\phi$.

Dipole truncations of the inner and outer general solutions for the vector field do not match
smoothly on the sphere at $r=r_0$. However, a combination of the inner and outer dipole truncations
of the loop solution will at least give a reasonable approximation away from $r=r_0$. For large
$r$, with zero net charge the leading terms in the vector potential are a combination of a
magnetic dipole and a charge quadrupole,
\begin{eqnarray}
A_t&\sim&{d\over r^3}P_2(\cos\theta)\\
A_\phi&\sim&{m\over r}\sin^2\theta
\end{eqnarray}
The dipole and quadrupole strenghts are
\begin{equation}
m=-\frac23(r_+-M)^2\beta_1^i,\quad d=\frac25(r_+-M)^3\beta_2^r-\frac23a(r_+-M)^2\beta_1^i.
\end{equation}
For the zero-EMF solution, these are given by
\begin{eqnarray}
m&=&{\pi I r_0\Delta(r_0)\over r_0-2M}\\
d&=&-{\pi I aM \Delta(r_0)\over r_0-2M}
\end{eqnarray}
For large loops with $r_0>>r_+$, these reduce to the ordinary magnetic dipole strength
$m=I\pi r_0^2$ and an electric quadropole moment $d=-\pi IaMr_0$.

\subsection{Extreme Kerr limit in the equatorial plane}

In the extremal Kerr limit the parameter $u\to\infty$ and the asymptotic forms of the
Legendre functions in the general loop solution can be used. The mode summations
can be done in a similar way to those of Sect. \ref{sec}, and the results expressed in terms
of hypergeometric functions,
\begin{equation}
F_c(r)=F\left(\frac12,\frac12;c;\left({r-M\over r_0-M}\right)^2\right).
\end{equation}
In the equatorial plane for $r<r_0$,
\begin{equation}
A_t=\left(1-{M\over r}\right)\left(\zeta(r_0)F_1(r)+\eta(r_0)F_2(r)\right)
+{2\pi IM^2\over r_0-2M}{1\over r}-{Q\over r},\label{ekat}
\end{equation}
where
\begin{eqnarray}
\zeta(r_0)&=&-{2\pi IM^2+q(r_0-2M)\over (r_0-M)^2},\label{zetadef}\\
\eta(r_0)&=&{2\pi IM^2(r_0+M)-qM^2\over 2 r_0 (r_0-M)^2}.\label{etadef}
\end{eqnarray}
The outer solution for $r>r_0$ is given by the same expression provided that values of
the hypergeometric function on the branch cut are defined by taking the average of values just
above and below the branch cut.  The zero-EMF solutions with vanishing total charge can be obtained
by fixing the black hole charge to $Q=-q$ and using Eq. (\ref{totq}),
\begin{equation}
A_t={\pi I M^2F_2(r)\over (r_0-2M)(r_0-M)}\left(1-{M\over r}\right)
-{2\pi IM^2\over r_0-2M}{1\over r}.
\end{equation}
This is finite at the ring, as we discussed in the previous section. The simple form of the solution
accounts for the high accuracy of the dipole approximation.

The axial part is not quite so simple. The axial vector potential is
\begin{equation}
A_\phi={C_1(r,r_0)\over r_0(r_0-M)^2}{F_1(r)\over r}
-{C_2(r,r_0)\over r_0(r_0-M)^2}\left(1-{M\over r}\right)F_2(r)
+{(2\pi IM-q)M\over r_0 r}-{QM\over r}+2\pi IM
\end{equation}
where
\begin{eqnarray}
C_1(r)&=&2\,\pi \,Ir_0 \left( r_0-M \right) {r}^{3}+ \left( 2\,\pi \,IM^2-2\,\pi \,I{r_0
}^{2}+4\,\pi IMr_0-qM \right)M {r}^{2}\nonumber\\
&&+2\,\pi \,I{r_0}^{2} \left( r_0-M
 \right)M r-4\pi I r_0M^4-q(r_0-2M)r_0M^2\\
C_2(r)&=&2\,\pi \,Ir_0 \left( r_0-1 \right) {r}^{2}+ \left( 2\,\pi \,IM^2-2\,\pi \,I{r_0
}^{2}+4\,\pi \,IMr_0 -qM\right)M r+2\,\pi \,I(r_0+M)M^3-qM^3
\end{eqnarray}

\subsection{Extreme Kerr with disks}\label{disks}

The availability of a closed form solution for equatorial current loops around
extreme Kerr black holes allows us to examine what happens with thin equatorial disks.
To construct models of this general type we must impose some physical condition
on the disc. A suitable choice would be for the electric field to vanish is some specially
chosen rotating frame with angular speed $\omega(r)$,
\begin{equation}
A_{t,r}+\omega A_{\phi,r}=0\label{dc}
\end{equation}
This allows us to solve for the charge density $q(r)$ given a current distribution $I(r)$.
as an example, we shall examine the simplest case, $\omega=0$, and find the charge distribution in a
disk which has zero potential in the Boyer-Lindquist frame.

We can obtain the solution inside the disc by combining the solutions (\ref{ekat}) for rings,
\begin{equation}
A_t=\left(1-{M\over r}\right)\int_{r_1}^\infty\left[\zeta(r')F_1(r)+\eta(r')F_2(r)\right]dr'
+{1\over r}\int_{r_1}^\infty{2\pi IM^2\over r_0-2M}dr'-{Q\over r},\label{disk}
\end{equation}
where $r_1$ is the inner edge of the ring. A zero potential solution can be obtained by having the
first integral vanish and by fixing the black hole's charge
\begin{equation}
Q=\int_{r_1}^\infty{2\pi IM^2\over r_0-2M}dr'.
\end{equation}
In order to find out when the first integral vanishes, we use the identity
$4F_1=(xF_2)'+2F_2$ to eliminate $F_1$. After integration by parts, this leads to two conditions,
\begin{equation}
\zeta(r_1)=0,\quad 4\zeta(r)+4\eta(r)+(r-M)\zeta(r)_{,r}=0.
\end{equation}
These can be regarded as equations which determine the charge density for a given
current density, by substituting Eqs. (\ref{zetadef}) and  (\ref{zetadef}). 
The first condition is the finite-potential condition,
\begin{equation}
q(r_1)=-{4\pi M^2 I(r_1)\over r_1-2M}.
\end{equation}
The second condition is a first order differential equation,
\begin{equation}
\left[r(r-2M)^2 q\right]_{,r}+M^2(r-2M)(4\pi rI)_{,r}=0.
\end{equation}
The solution is
\begin{equation}
q(r)=-{4\pi M^2 I(r)\over r-2M}+{4\pi M^2\over r(r-2M)^2}\int_{r_1}^rI(r') r'dr'.
\end{equation}
This is only valid for $r_1>2M$. Unfortunately, this appears to be the only case of (\ref{dc}) which
seems solvable analytically.

\section{Conclusion}

The main result of this paper has been to correct the published solutions to Maxwell's equations for
the electromagnetic vector potential near to a Kerr black hole surrounded by a charged current
loop. The vector potential is a necesary ingredient for the determination of charged particle
orbits close to a black hole immersed in a magnetic field. These orbits can be used to determine
the inner edge of an accretion disc. They could also be used to extend recent work on energy flux
generation using the Penrose pair creation $\gamma\gamma\to e^+e^-$ and Penrose Compton scattering 
\cite{Williams:2004}  to include magnetic fields.

Current loops can be superposed to form non-vacuum solutions \cite{Li:2000}. In these situations
the loops will be charged.  For example, an equatorial conducting disk with angular speed
$\omega_D$ would have a zero charge density in the disk frame, but a non-zero charge
density in the Boyer-Lindquist coordinate frame. This can be modeled by a superposition of charged
current loops in a similar way  to the analysis in Sect \ref{disks}. It may also be of interest to
combine the fields due to a current along the rotation axis with the current rings to produce a toy
model which has a current circulating out along the equatorial plane
and back down the rotation axis.

\appendix

\bibliography{paper.bib}

\begin{thebibliography}{29}
\expandafter\ifx\csname natexlab\endcsname\relax\def\natexlab#1{#1}\fi
\expandafter\ifx\csname bibnamefont\endcsname\relax
  \def\bibnamefont#1{#1}\fi
\expandafter\ifx\csname bibfnamefont\endcsname\relax
  \def\bibfnamefont#1{#1}\fi
\expandafter\ifx\csname citenamefont\endcsname\relax
  \def\citenamefont#1{#1}\fi
\expandafter\ifx\csname url\endcsname\relax
  \def\url#1{\texttt{#1}}\fi
\expandafter\ifx\csname urlprefix\endcsname\relax\def\urlprefix{URL }\fi
\providecommand{\bibinfo}[2]{#2}
\providecommand{\eprint}[2][]{\url{#2}}

\bibitem[{\citenamefont{Punsly}(2001)}]{Punsley:2001}
\bibinfo{author}{\bibfnamefont{B.}~\bibnamefont{Punsly}},
  \emph{\bibinfo{title}{{Black Hole Gravitohydromagnetics}}}
  (\bibinfo{publisher}{Springer, Berlin}, \bibinfo{year}{2001}).

\bibitem[{\citenamefont{Thorne et~al.}(1986)\citenamefont{Thorne, Price, and
  Macdonald}}]{Thorne:1986iy}
\bibinfo{author}{\bibfnamefont{K.~S.} \bibnamefont{Thorne},
  \bibfnamefont{(Ed.~)}}, \bibinfo{author}{\bibfnamefont{R.~H.}
  \bibnamefont{Price}, \bibfnamefont{(Ed.~)}}, \bibnamefont{and}
  \bibinfo{author}{\bibfnamefont{D.~A.} \bibnamefont{Macdonald},
  \bibfnamefont{(Ed.~)}}, \emph{\bibinfo{title}{{Black holes: the membrane
  paradigm}}} (\bibinfo{publisher}{New Haven, USA: Yale Univ. Pr.},
  \bibinfo{year}{1986}).

\bibitem[{\citenamefont{Blandford and Znajek}(1977)}]{Blandford:1977}
\bibinfo{author}{\bibfnamefont{R.}~\bibnamefont{Blandford}} \bibnamefont{and}
  \bibinfo{author}{\bibfnamefont{R.}~\bibnamefont{Znajek}},
  \bibinfo{journal}{Mon. Not. Roy. Astron. Soc.}
  \textbf{\bibinfo{volume}{179}}, \bibinfo{pages}{433} (\bibinfo{year}{1977}).

\bibitem[{\citenamefont{Aliev and Ozdemir}(2002)}]{Aliev:2002nw}
\bibinfo{author}{\bibfnamefont{A.~N.} \bibnamefont{Aliev}} \bibnamefont{and}
  \bibinfo{author}{\bibfnamefont{N.}~\bibnamefont{Ozdemir}},
  \bibinfo{journal}{Mon. Not. Roy. Astron. Soc.}
  \textbf{\bibinfo{volume}{336}}, \bibinfo{pages}{241} (\bibinfo{year}{2002}),
  \eprint{gr-qc/0208025}.

\bibitem[{\citenamefont{Prasanna}(1978)}]{Prasanna:1978}
\bibinfo{author}{\bibfnamefont{C.}~\bibnamefont{Prasanna}, \bibfnamefont{A.R.
  amd~Vishveshwara}}, \bibinfo{journal}{Pramana} \textbf{\bibinfo{volume}{11}},
  \bibinfo{pages}{359} (\bibinfo{year}{1978}).

\bibitem[{\citenamefont{Prasanna}(1980)}]{Prasanna:1980}
\bibinfo{author}{\bibfnamefont{C.}~\bibnamefont{Prasanna}, \bibfnamefont{A.R.
  amd~Vishveshwara}}, \bibinfo{journal}{Riv. Nuovo Cimento}
  \textbf{\bibinfo{volume}{I1}}, \bibinfo{pages}{1} (\bibinfo{year}{1980}).

\bibitem[{\citenamefont{Sengupta}(1997)}]{Sengupta:1997na}
\bibinfo{author}{\bibfnamefont{S.}~\bibnamefont{Sengupta}},
  \bibinfo{journal}{Int. J. Mod. Phys.} \textbf{\bibinfo{volume}{D6}},
  \bibinfo{pages}{591} (\bibinfo{year}{1997}), \eprint{gr-qc/9707014}.

\bibitem[{\citenamefont{Takahashi and Koyama}(2009)}]{Takahashi:2008zh}
\bibinfo{author}{\bibfnamefont{M.}~\bibnamefont{Takahashi}} \bibnamefont{and}
  \bibinfo{author}{\bibfnamefont{H.}~\bibnamefont{Koyama}},
  \bibinfo{journal}{Astrophys. J.} \textbf{\bibinfo{volume}{693}},
  \bibinfo{pages}{472} (\bibinfo{year}{2009}).

\bibitem[{\citenamefont{Wagh and Dadhich}(1985)}]{Wagh:1985}
\bibinfo{author}{\bibfnamefont{S.}~\bibnamefont{Wagh},
  \bibfnamefont{S.M.~Dhurandhar}} \bibnamefont{and}
  \bibinfo{author}{\bibfnamefont{N.}~\bibnamefont{Dadhich}},
  \bibinfo{journal}{Astrophys.J.} \textbf{\bibinfo{volume}{290}},
  \bibinfo{pages}{12} (\bibinfo{year}{1985}).

\bibitem[{\citenamefont{Bardeen and Teukolsky}(1972)}]{Bardeen:1972}
\bibinfo{author}{\bibfnamefont{W.}~\bibnamefont{Bardeen},
  \bibfnamefont{J.M.~Press}} \bibnamefont{and}
  \bibinfo{author}{\bibfnamefont{S.}~\bibnamefont{Teukolsky}},
  \bibinfo{journal}{Astrophys.J.} \textbf{\bibinfo{volume}{178}},
  \bibinfo{pages}{347} (\bibinfo{year}{1972}).

\bibitem[{\citenamefont{Wald}(1974{\natexlab{a}})}]{Wald:1974b}
\bibinfo{author}{\bibfnamefont{R.}~\bibnamefont{Wald}},
  \bibinfo{journal}{Astrophys.J.} \textbf{\bibinfo{volume}{191}},
  \bibinfo{pages}{231} (\bibinfo{year}{1974}{\natexlab{a}}).

\bibitem[{\citenamefont{Wagh and Dadhich}(1989)}]{Wagh:1989}
\bibinfo{author}{\bibfnamefont{S.}~\bibnamefont{Wagh}} \bibnamefont{and}
  \bibinfo{author}{\bibfnamefont{N.}~\bibnamefont{Dadhich}},
  \bibinfo{journal}{Phys. Rep.} \textbf{\bibinfo{volume}{183}},
  \bibinfo{pages}{137} (\bibinfo{year}{1989}).

\bibitem[{\citenamefont{Chitre and Vishveshwara}(1975)}]{Chitre:1975}
\bibinfo{author}{\bibfnamefont{D.}~\bibnamefont{Chitre}} \bibnamefont{and}
  \bibinfo{author}{\bibfnamefont{C.}~\bibnamefont{Vishveshwara}},
  \bibinfo{journal}{Phys. Rev. D} \textbf{\bibinfo{volume}{12}},
  \bibinfo{pages}{1538} (\bibinfo{year}{1975}).

\bibitem[{\citenamefont{Petterson}(1975)}]{Petterson:1975}
\bibinfo{author}{\bibfnamefont{J.}~\bibnamefont{Petterson}},
  \bibinfo{journal}{Phys. rev. D} \textbf{\bibinfo{volume}{12}},
  \bibinfo{pages}{2218} (\bibinfo{year}{1975}).

\bibitem[{\citenamefont{Bi\v{c}{}\'ak and Dvo\v{r}{}\'ak}(1976)}]{Bicak:1976}
\bibinfo{author}{\bibfnamefont{J.}~\bibnamefont{Bi\v{c}{}\'ak}}
  \bibnamefont{and}
  \bibinfo{author}{\bibfnamefont{L.}~\bibnamefont{Dvo\v{r}{}\'ak}},
  \bibinfo{journal}{General Relativity and Gravitation}
  \textbf{\bibinfo{volume}{7}}, \bibinfo{pages}{959} (\bibinfo{year}{1976}).

\bibitem[{\citenamefont{Li}(2000)}]{Li:2000}
\bibinfo{author}{\bibfnamefont{L.-X.} \bibnamefont{Li}},
  \bibinfo{journal}{Phys. Rev. D} \textbf{\bibinfo{volume}{61}},
  \bibinfo{pages}{084016} (\bibinfo{year}{2000}).

\bibitem[{\citenamefont{{Punsly}}(1991)}]{Punsly:1981}
\bibinfo{author}{\bibfnamefont{B.}~\bibnamefont{{Punsly}}},
  \bibinfo{journal}{Astrophys. J.} \textbf{\bibinfo{volume}{372}},
  \bibinfo{pages}{424} (\bibinfo{year}{1991}).

\bibitem[{\citenamefont{Balbus and Hawley}(1991)}]{Balbus:1991ay}
\bibinfo{author}{\bibfnamefont{S.~A.} \bibnamefont{Balbus}} \bibnamefont{and}
  \bibinfo{author}{\bibfnamefont{J.~F.} \bibnamefont{Hawley}},
  \bibinfo{journal}{Astrophys. J.} \textbf{\bibinfo{volume}{376}},
  \bibinfo{pages}{214} (\bibinfo{year}{1991}).

\bibitem[{\citenamefont{Li}(2002)}]{Li:2000fn}
\bibinfo{author}{\bibfnamefont{L.-X.} \bibnamefont{Li}},
  \bibinfo{journal}{Astrophys. J.} \textbf{\bibinfo{volume}{567}},
  \bibinfo{pages}{463} (\bibinfo{year}{2002}), \eprint{astro-ph/0012469}.

\bibitem[{\citenamefont{Chandrasekhar}(1992)}]{Chandra:1992}
\bibinfo{author}{\bibfnamefont{S.}~\bibnamefont{Chandrasekhar}},
  \emph{\bibinfo{title}{{The mathematical theory of black holes}}}
  (\bibinfo{publisher}{Oxford, UK}, \bibinfo{year}{1992}).

\bibitem[{\citenamefont{Petterson}(1974)}]{Petterson:1974}
\bibinfo{author}{\bibfnamefont{J.}~\bibnamefont{Petterson}},
  \bibinfo{journal}{Phys. rev. D} \textbf{\bibinfo{volume}{10}},
  \bibinfo{pages}{3166} (\bibinfo{year}{1974}).

\bibitem[{\citenamefont{Chamblin et~al.}(1998)\citenamefont{Chamblin, Emparan,
  and Gibbons}}]{Chamblin:1998qm}
\bibinfo{author}{\bibfnamefont{A.}~\bibnamefont{Chamblin}},
  \bibinfo{author}{\bibfnamefont{R.}~\bibnamefont{Emparan}}, \bibnamefont{and}
  \bibinfo{author}{\bibfnamefont{G.~W.} \bibnamefont{Gibbons}},
  \bibinfo{journal}{Phys. Rev.} \textbf{\bibinfo{volume}{D58}},
  \bibinfo{pages}{084009} (\bibinfo{year}{1998}), \eprint{hep-th/9806017}.

\bibitem[{\citenamefont{Bi\v{c}{}\'ak and Dvo\v{r}{}\'ak}(1980)}]{Bicak:1980}
\bibinfo{author}{\bibfnamefont{J.}~\bibnamefont{Bi\v{c}{}\'ak}}
  \bibnamefont{and}
  \bibinfo{author}{\bibfnamefont{L.}~\bibnamefont{Dvo\v{r}{}\'ak}},
  \bibinfo{journal}{Phys. Rev. D} \textbf{\bibinfo{volume}{22}},
  \bibinfo{pages}{2933} (\bibinfo{year}{1980}).

\bibitem[{\citenamefont{Bi\v{c}{}\'ak and Janis}(1985)}]{Bicak:1985}
\bibinfo{author}{\bibfnamefont{J.}~\bibnamefont{Bi\v{c}{}\'ak}}
  \bibnamefont{and} \bibinfo{author}{\bibfnamefont{V.}~\bibnamefont{Janis}},
  \bibinfo{journal}{MNRAS} \textbf{\bibinfo{volume}{212}}, \bibinfo{pages}{899}
  (\bibinfo{year}{1985}).

\bibitem[{\citenamefont{Wald}(1974{\natexlab{b}})}]{Wald:1974}
\bibinfo{author}{\bibfnamefont{R.}~\bibnamefont{Wald}}, \bibinfo{journal}{Phys.
  rev. D} \textbf{\bibinfo{volume}{10}}, \bibinfo{pages}{1680}
  (\bibinfo{year}{1974}{\natexlab{b}}).

\bibitem[{\citenamefont{Linet}(1977)}]{Linet:1977}
\bibinfo{author}{\bibfnamefont{B.}~\bibnamefont{Linet}}, \bibinfo{journal}{C.R.
  Acad. Sci. Paris A} \textbf{\bibinfo{volume}{284}}, \bibinfo{pages}{1167}
  (\bibinfo{year}{1977}).

\bibitem[{\citenamefont{Linet}(1979)}]{Linet:1979}
\bibinfo{author}{\bibfnamefont{B.}~\bibnamefont{Linet}},
  \bibinfo{journal}{Journal of Physics A: Mathematical and General}
  \textbf{\bibinfo{volume}{12}}, \bibinfo{pages}{839} (\bibinfo{year}{1979}).

\bibitem[{\citenamefont{Aliev and Galtsov}(1988)}]{Aliev:1988wy}
\bibinfo{author}{\bibfnamefont{A.~N.} \bibnamefont{Aliev}} \bibnamefont{and}
  \bibinfo{author}{\bibfnamefont{D.~V.} \bibnamefont{Galtsov}},
  \bibinfo{journal}{Sov. Phys. JETP} \textbf{\bibinfo{volume}{67}},
  \bibinfo{pages}{1525} (\bibinfo{year}{1988}).

\bibitem[{\citenamefont{Williams}(2004)}]{Williams:2004}
\bibinfo{author}{\bibfnamefont{R.~K.} \bibnamefont{Williams}},
  \bibinfo{journal}{The Astrophysical Journal} \textbf{\bibinfo{volume}{611}},
  \bibinfo{pages}{952} (\bibinfo{year}{2004}).

\end{thebibliography}

\end{document}